\documentclass[fleqn,useAMS,usenatbib,usegraphicx]{mn2e}
\usepackage{times}
\usepackage{graphics,graphicx}
\usepackage{natbib}
\usepackage{epsfig}
\usepackage{rotating}
\usepackage{lscape}
\usepackage{amssymb}

\begin{document}

\newcommand{\hi}{H{\sc i}}
\newcommand{\nhi}{{\rm N_{\rm HI}}}
\newcommand{\hii}{H{\sc i}\,21cm}
\newcommand{\htwo}{{\rm H}_2}
\newcommand{\al}{\ensuremath{\alpha}}
\newcommand{\dal}{\ensuremath{\lsb \Delta \alpha/ \alpha \rsb}}
\newcommand{\gp}{\ensuremath{{g}_{p}}}
\newcommand{\lqcd}{\ensuremath{\Lambda_{\rm QCD}}}
\newcommand{\y}{\ensuremath{{m}_{e}/{m}_{p}}}
\newcommand{\dmu}{\ensuremath{\lsb \Delta \mu/\mu \rsb}}
\newcommand{\beq}{\begin{equation}}
\newcommand{\eeq}{\end{equation}}
\newcommand{\noi}{\noindent}
\newcommand{\lb}{\left(}
\newcommand{\cm}{cm$^{-2}$}
\newcommand{\rb}{\right)}
\newcommand{\lsb}{\left[}
\newcommand{\rsb}{\right]}
\newcommand{\kms}{km~s$^{-1}$}
\newcommand{\msun}{M$_\odot$}
\newcommand{\mjyb}{\ensuremath{{\rm mJy \, bm}^{-1}}}
\newcommand{\mb}{M\ensuremath{_B}}
\def\aa{A\&A}
\def\mnras{MNRAS}
\def\na{New Astronomy}
\def\apj{ApJ}
\def\memras{MmRAS}
\def\aaps{A\&AS}
\def\aap{A\&A}
\def\apjs{ApJS}
\def\apjl{ApJL}
\def\aj{AJ}
\def\aar{A\&AR}
\def\pasj{PASJ}
\def\araa{ARA\&A}
\def\apss{Ap\&SS}
\def\degree{$^{\circ}$}
\def\schi{{\sc Hi}\ }
\def\ga{\mathrel{\hbox{\rlap{\hbox{\lower4pt\hbox{$\sim$}}}\hbox{$>$}}}}
\def\la{\mathrel{\hbox{\rlap{\hbox{\lower4pt\hbox{$\sim$}}}\hbox{$<$}}}}

\pagestyle{myheadings}

\title[H{\sc i} 21cm emission from a sub-DLA at $z = 0.0063$]
{H{\sc i} 21cm  emission from the sub-damped Lyman-$\alpha$ absorber at 
$z = 0.0063$ towards PG\,1216+069}

\author[Chengalur et al.]{J. N. Chengalur$^1$\thanks{E-mail: chengalu@ncra.tifr.res.in},
T. Ghosh$^2$, 
C. J. Salter$^2$, 
N. Kanekar$^1$,
E. Momjian$^3$,
B.A. Keeney$^4$,
\newauthor
J.T. Stocke$^4$\\
$^1$National Centre for Radio Astrophysics, TIFR, Ganeshkhind, Pune - 411007, India\\
$^2$NAIC, Arecibo Observatory, HC3 Box 53995, Arecibo, PR 00612\\
$^3$National Radio Astronomy Observatory, 1003 Lopezville Road, Socorro, NM 87801-0387 \\
$^4$Center for Astrophysics and Space Astronomy, Department of Astrophysical and 
Planetary Sciences, Box 389, University of Colorado, Boulder, CO 80309
}
\maketitle

\begin{abstract}

We present H{\sc i}\,21cm emission observations of the $z \sim 0.00632$
sub-damped  Lyman-$\alpha$ absorber (sub-DLA) towards PG\,1216+069 made using
the Arecibo Telescope and the Very Large Array (VLA). The Arecibo H{\sc i}\,21cm 
spectrum corresponds to an H{\sc i} mass of $\sim 3.2\times 10^7$~M$_\odot$, 
two orders of magnitude smaller than that of a typical spiral galaxy. This
is surprising since in the local Universe the cross-section for absorption
at high H{\sc i} column densities is expected to be dominated by spirals.
The  H{\sc i}\,21cm emission detected in the VLA spectral cube has a low
signal-to-noise ratio, and represents only half the total flux seen at 
Arecibo. Emission from three other sources is detected in the VLA observations, 
with only one of these sources having an optical counterpart. This group of 
H{\sc i} sources appears to be part of complex ``W'', believed to lie in 
the background of the Virgo cluster. While several H{\sc i} cloud complexes have 
been found in and around the Virgo cluster, it is unclear whether the ram pressure and
galaxy harassment processes that are believed to be responsible for
the creation of such clouds in a cluster environment are relevant at
the location of this cloud complex.  The extremely low metallicity of the
gas, $\sim 1/40$-solar, also makes it unlikely that the sub-DLA consists
of material that has been stripped from a galaxy. Thus, while our results
have significantly improved our understanding of the host of this sub-DLA, 
the origin of the gas cloud remains a mystery.

\end{abstract}


\section{Introduction}
\label{sec1}

Much of what we know about the interstellar medium of high redshift galaxies
comes from studies of absorption lines seen in the spectra of distant quasars. 
In this context, the most interesting systems are those with the highest neutral
hydrogen (\hi) column densities: in the local Universe, such high column density gas 
is almost invariably associated with galaxies. The highest \hi\ column density systems are
also interesting because the bulk of neutral atomic gas at high redshifts 
is provided by absorbers with \hi\ column densities $\geq 2 \times 10^{20}$~cm$^{-2}$, 
the so-called damped Lyman-$\alpha$ absorbers \citep[DLAs;][]{wolfe05}. At these
column densities, the gas is predominantly neutral, making DLAs the best known 
reservoirs for star formation at high redshifts. Absorption systems with 
slightly lower \hi\ column densities (in the range $10^{19} - 2 \times 
10^{20}$~\cm, the ``sub-DLAs'') are significantly ionised, but nonetheless
contribute a non-trivial fraction of the total neutral gas at high 
redshifts. Sub-DLAs are also expected to be arise in galaxies \citep[e.g.][]{peroux05}, 
possibly in their outer regions.

Since absorption-selected galaxy samples contain no luminosity bias, studies of 
such samples provide information on the nature of ``typical'' galaxies at 
different redshifts. However, despite over two decades of study, 
\citep[e.g.][]{moller93,prochaska97, haehnelt98,warren01,kulkarni10,fynbo10,fynbo11,fumagalli15}, 
the nature of the host galaxies of DLAs and sub-DLAs remains unclear.
On the theoretical front, numerical simulations \citep[e.g.][]{pontzen08,
rahmati14} find that DLA hosts span a range of halo masses, albeit with 
a peak at low mass. For example, \citet{rahmati14} find that most DLAs are 
associated with low-mass ($\lesssim 10^{10}$~\msun) halos, but that the 
highest \hi\ column density DLAs arise in more massive galaxies. 

Unfortunately, there are very few direct observational constraints on these 
theoretical models for the host galaxies of high redshift DLAs. This is 
because identification of the host galaxies of  DLAs at $z \gtrsim 2$ via 
direct imaging has proved to be extremely difficult, even with the Hubble 
Space Telescope (HST) or 8m-class ground-based optical telescopes. The faint 
stellar emission from the foreground galaxy typically lies directly below 
the bright emission from the background quasar making it very difficult
to identify the host. Galaxy counterparts are hence known only for a handful of 
high-$z$ DLAs, mostly systems with high metallicities (e.g. \citealp{fynbo10,
fynbo11, krogager12}; but see also \citealp{noterdaeme12}).  This dynamic 
range problem can be overcome for sightlines with at least two high column 
density absorbers -- the higher-redshift absorber can be used as a filter to block 
the quasar radiation, allowing a clean measurement of the stellar emission from 
the lower redshift system \citep{omeara06,fumagalli10,fumagalli14b}. 
Observations of 32 DLAs using this technique have led to stringent 
constraints on the star formation rate (SFR) of their host galaxies, with 
the 2$\sigma$ SFR limits of $\approx 0.09 - 0.27 M_\odot $~per year
suggesting that the hosts are small galaxies with low star formation 
\citep{fumagalli14b,fumagalli15}. This picture is consistent with both 
the typical low metallicities of the absorbers \citep[e.g.][]{pettini94,prochaska03,kulkarni05},
as well as with their typical high spin temperatures \citep[e.g.][]{kanekar03,kanekar14}.

At low redshifts ($z \lesssim 1$), information is now available 
on the optical luminosity, stellar mass, star formation rate, etc. of 
about twenty DLAs and sub-DLAs \citep[e.g.][]{lebrun97,rao03,rao11,chen05}. 
However, little is known about their gas mass and distribution. This is 
unfortunate, since the gas distribution is a critical physical parameter 
in absorption-selected samples. If one extrapolates from the properties of 
the $z\sim 0$ galaxy population, one would expect the cross-section for 
DLA absorption, particularly at the high \hi\ column density end, to be 
dominated by large spiral galaxies \citep{zwaan05,patra13}. \hii\ emission 
studies of DLAs allow a direct determination of the gas mass, the velocity 
field, and the size of the gas reservoir of the host system. Unfortunately,
the weakness of the \hii\ transition has meant that such searches have only 
been possible in a handful of DLAs and sub-DLAs, those at the lowest redshifts,
$z \lesssim 0.1$ \citep[][]{kanekar01,bowen01b,chengalur02,kanekar05c,
mazumdar14}. To date, good quality \hii\ imaging is available for only one 
DLA, at $z = 0.009$ towards SBS~1543+593 \citep{bowen01b,chengalur02}. A 
tentative detection of \hii\ emission has been reported in a sub-DLA at 
$z = 0.006$ towards PG\,1216+069 \citep{briggs06,tripp08}, and strong upper 
limits on the \hi\ mass, $\lesssim 3 \times 10^9 M_\odot $, have been 
obtained for three other DLAs and sub-DLAs, all at $z \approx 0.1$ 
\citep{kanekar01, mazumdar14}. 

In this paper, we present results from a search for \hii\ emission from 
the $z = 0.00632$ sub-DLA towards PG\,1216+069. This system was observed
using the Space Telescope Imaging Spectrograph (STIS) onboard the HST 
by \citet{tripp05} as part of a large survey for low-redshift O{\sc vi} 
absorbers, but was discovered serendipitously to have a damped Ly-$\alpha$
profile. Follow-up \hii\ emission observations were conducted by a number
of authors, with tentative detections of \hii\ emission reported using the 
Westerbork Synthesis Radio Telescope \citep[WSRT;][]{briggs06} and the Very 
Large Array \citep[VLA;][]{tripp08}. Here, we present results based on observations 
with the Arecibo 305-m telescope, as well as a re-analysis of the archival 
VLA data. The Arecibo spectrum has an excellent signal-to-noise ratio, allowing 
us to robustly measure the \hi\ content of the galaxy. The VLA data allow us 
to further constrain the spatial extent of the emitting gas, as well as to 
study the environment  of the host system. The rest of this paper is arranged 
as follows. Section~\ref{sec2} presents the observations and data reduction,
while our results are presented in Section~\ref{sec3} and discussed in Section~\ref{sec4}. 

\section{Observations and Data Reduction}
\label{sec2}

The Arecibo telescope observations (centred at RA$_{\rm J2000}$=12:19:20.88,
DEC$_{\rm J2000}$=+06:38:38.4) of the $z=0.00632$ sub-DLA towards PG\,1216+069 were
split into a number of short runs between 2004 December and 2005 April, using 
the L-Band Wide (LBW) receiver and two orthogonal linear polarisations. Passbands 
of 12.5 and 25~MHz centred on the redshifted \hii\ line frequency for 
$z =0.00632$ were observed simultaneously, each sub-divided into 2048 frequency
channels by the spectrometer. Standard position-switching (ON/OFF) was used for
bandpass calibration. For a single scan, both the ON and the OFF phases had a 
duration of 4~min, with each ON/OFF pair followed by a noise diode calibration
of the brightness scale. Spectra were written to disk every 1~sec for 
subsequent analysis.  

The data were  analysed using the Arecibo IDL reduction package of Phil Perillat. At the 
time of the observations, L-band measurements at Arecibo were suffering from a locally generated 
interference that wandered over the spectrum on a time scale of minutes, contaminating a 
significant fraction of the data. An IDL subroutine was written to eliminate the worst effects 
of this interference.  This was applied to the data of all scans before the quantity 
(ON$-$OFF)/OFF was calculated, and the spectra calibrated into units of Jy\,beam$^{-1}$. 
The interference excision procedure succeeded in salvaging the vast majority of affected 
scans, and very few had to be rejected. A weighted average of the individual Hanning-smoothed 
spectra was then taken and the two polarisations averaged to produce the final spectrum.

After a clear Arecibo detection of \hii\ emission from the sub-DLA had been obtained, 
data were acquired using the same ON/OFF mode for four positions offset celestially 
north, south, east and west by 1.6~arcmin, i.e. roughly half the telescope half-power beamwidth 
(HPBW), from the position of PG\,1216+069. This was done to investigate whether the \hii\ emission 
from the sub-DLA (a)~was extended, (b)~was offset relative to PG\,1216+069, or (c)~showed 
kinematical signatures of rotation. These data were analysed in a similar manner to 
the original data.

The VLA data were obtained in project AT312 (PI: Tripp) on 2005 December 22, when the array was 
in its 'D' configuration \citep{tripp08}. The total on-source time was $\sim7$~hr. The 
observations used a bandwidth of 3.125~MHz ($\sim 660$~\kms), divided into 64 spectral 
channels, giving a channel width of 10.4~\kms. The integrated \hii\ emission obtained 
from these data was earlier presented by \citet{tripp08}. We re-analysed these data,
after downloading them from the VLA archive, using standard procedures in classic {\sc aips}. 
After initial flux density, gain and bandpass calibration, the visibilities of the target 
source were separated out for self-calibration. After a single round of phase-only 
self-calibration, a strong source at the edge of the field of view was subtracted out using 
the task {\sc uvsub}. Following this, the field was imaged again and the brightest detected 
continuum sources subtracted out again using {\sc uvsub}. The field was then imaged once 
more, and the residual continuum subtracted using the task {\sc imlin}. The 
continuum emission from the quasar PG\,1216+069 itself is fairly weak 
\citep[$\sim few$~mJy;][]{kellerman89,kanekar05c}, and does not pose any problems when 
trying to measure the \hi\ emission along this line of sight. Spectra for all detected objects 
were finally extracted from the spectral cube, after correcting for the shape of the 
VLA primary beam, using the task {\sc blsum}.

\section{Results}
\label{sec3}
\begin{table*}
\caption{Results from the five Arecibo pointings around PG\,1216+069.}
\begin{center}
\begin{tabular}{lccccccc}
\hline
Posn &  T$_{int}$ & RMS noise & S$_{peak}$ &  S$_{int}$ &  Int$_{\rm{norm}}$  &
V$_{{\rm peak}}$  &    w$_{\rm{50}}$ \\
 &       (min) &  (mJy\,bm$^{-1}$) & (mJy\,bm$^{-1}$) & (Jy\,km\,s$^{-1}$)
& & km\,s$^{-1}$   &      km\,s$^{-1}$ \\
\hline
Center &  56 &   0.16 &   2.56 &   0.1781  & -- &  1914 & 68 \\
North &   36 &   0.19 &   1.51 &   0.1013  &  0.579 & 1914 & 58 \\
South &   28 &   0.20 &   1.38 &   0.0901  &  0.523 & 1916 & 64 \\
East &    20 &   0.24 &   1.12 &   0.1030  &   0.508 & 1905 & 95 \\
West &    16 &   0.25 &   1.41 &   0.0867  &   0.519 & 1913 & 46 \\
\hline
\end{tabular}
\end{center}
\label{5pt}
\end{table*}

\setcounter{figure}{0}
\begin{figure}
\centering
\includegraphics[width=8.5cm]{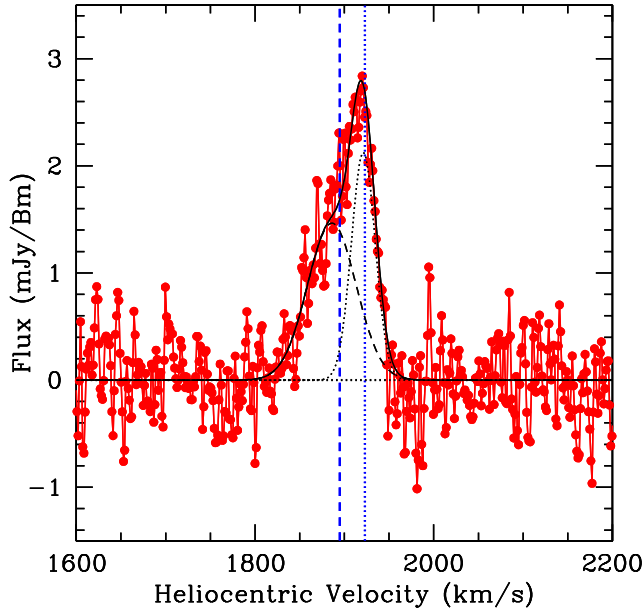}
\caption{The Arecibo \hii\ emission spectrum from the $z = 0.00632$ sub-DLA 
towards PG\,1216+069, at a velocity resolution of $\approx 2.6$~\kms.  The RMS 
noise on the spectrum is 0.35~$\mjyb$. The dashed vertical line indicates the 
redshift measured from the metal lines (viz. $z = 0.00632$), while the dotted 
line indicates the velocity of the weaker absorption seen in the metal lines 
by \citet{tripp05}. The error in the metal line  absorption velocity scale 
is $5-10$~\kms\ \citep{tripp05}. A double Gaussian fit to the Arecibo \hi\ 
profile is also shown. As can be seen, the velocity of the narrow \hii\ component 
matches that of the weaker metal-line absorption.
}
\label{fig1}
\end{figure}

Figure~\ref{fig1} shows the final Arecibo \hii\ spectrum from the 12.5-MHz passband, 
at a velocity resolution of $\approx 2.6$~\kms; this is based on 56~min of ON-source 
data and has a root-mean-square (RMS) noise of 0.35~$\mjyb$. \hii\ emission from the 
sub-DLA is detected at the expected redshifted line frequency with a signal-to-noise 
ratio (S/N) of $\approx 8$ per independent channel. The position and velocity of this 
object indicates that it is a member of cloud ``W'' which lies in the background of the 
Virgo cluster \citep{devaucouleurs61,mei07}, and is believed to be at about twice the 
Virgo cluster distance \citep{binggeli93}. The systemic redshift of $z=0.00632$  
corresponds to a distance  of $\approx 28$~Mpc (i.e. approximately twice the distance 
to the Virgo cluster) in a Lambda cold dark matter (LCDM)  cosmology with 
H$_0 =68$~\kms\,Mpc$^{-1}$, $\Omega_{M} = 0.31$,  $\Omega_{\rm{\Lambda}} = 0.69$. 
We use this distance to compute all distance-dependent quantities in this paper.  
At this assumed distance, the integrated flux density corresponds to an \hi\ mass 
of $\sim 3.2\times10^7$~\msun.

\begin{figure}
\includegraphics[width=9.0cm]{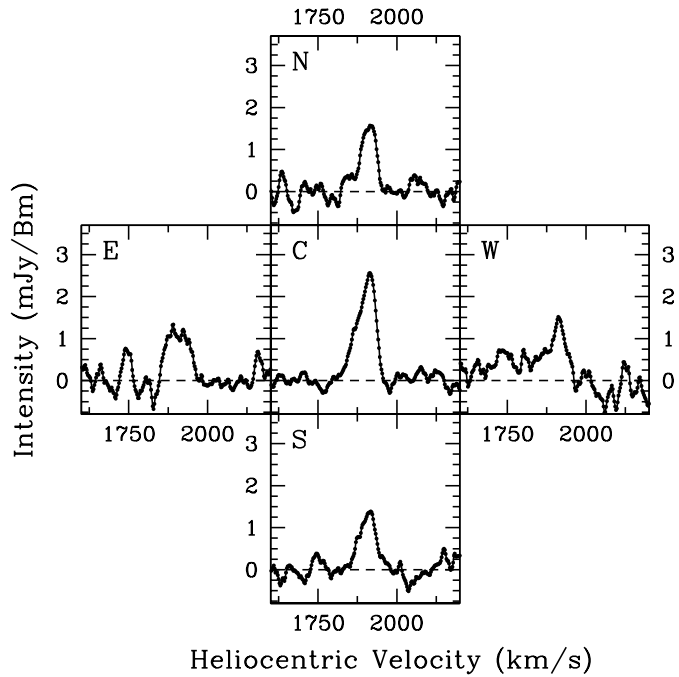}
\caption{\hii\ emission spectra derived from the mosaic of pointings around the 
position of PG\,1216+069. The displayed spectra are towards the quasar location, 
and positions offset to the celestial north, east, south and west by 1.6~arcmin 
(i.e. $\approx$ half of the telescope HPBW). The observations indicate that the 
\hii\ emission detected at Arecibo is unresolved. See the main text for more 
details.
}
\label{fig:amosaic}
\end{figure}

The Arecibo \hii\ spectra from the original and the offset-position observations  
are  shown in Fig.~\ref{fig:amosaic}. The spectra have been smoothed to  a velocity 
resolution of 24~\kms\ to maximise the S/N. At this velocity resolution, the peak 
S/N at the central location is $\approx 16$ per independent velocity channel. The 
observational and derived parameters for each Arecibo pointing position are 
presented in Table~\ref{5pt}. The columns are (1)~the offset direction from 
PG\,1216+069,  (2)~the ON-source integration time, (3)~the RMS noise, (4)~the peak 
\hii\ flux density, (5)~the \hii\ line integral, in Jy~\kms, (6)~the average of the 
peak \hii\ intensity and the \hii\ line integral at an offset position after 
normalisation by the corresponding values for the central position, (7)~the 
heliocentric radial velocity at peak \hii\  line intensity, and (8)~the 
velocity width of the \hii\ line at half power, all at a velocity 
resolution of 24~\kms.  

\begin{figure}
\includegraphics[width=7.0cm,angle=0]{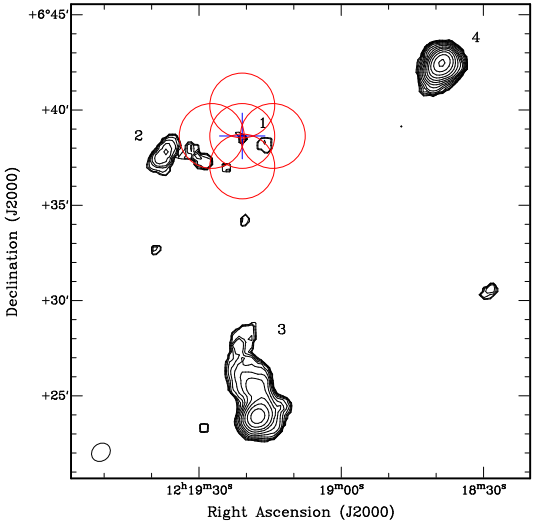}
\caption{The integrated \hii\ emission detected in the VLA data cube with
a resolution of $63^{''}\times 53^{''}$; the first 
contour is at an \hi\ column density of $2 \times 10^{18}$~\cm, with the \hi\ 
column density of successive contours increasing in steps of $\sqrt{2}$. 
The cross marks the location of PG\,1216+069, while the circles indicate the 
HPBW of the Arecibo pointings, resulting in the spectra shown in 
Figure~\ref{fig:amosaic}. The four sources detected in the VLA cube are numbered,
with numbers increasing anticlockwise about the centre. Source~4 corresponds to the 
galaxy VCC\,297, while the other sources are not associated with any known 
galaxy.
}
\label{fig:mom0a}
\end{figure}

\hii\ emission from four separate sources was detected in the VLA data cube; 
the integrated \hii\ emission from these sources is shown in Fig.~\ref{fig:mom0a}. 
The cross indicates the position of the background quasar PG\,1216+069, while the 
circles represent the HPBW of the different Arecibo pointings. The position of one 
of the \hii\ sources coincides with the quasar location; this source is likely to be 
the host of the sub-DLA. The derived parameters of the 
different sources are listed in Table~\ref{tab:vflux}, whose columns are
(1)~the source number, as given in Fig.~\ref{fig:mom0a}, (2)~the central velocity 
(as estimated from a single-Gaussian fit to the spectral profile), (3)~the 50\% 
velocity width (as estimated from the above fit), (4)~the integrated \hii\ flux 
density (Jy~\kms), and (5)~the \hi\ mass of the gas cloud.

\begin{table*}
\caption{Parameters of the \hii\ emission detected in the VLA observations}
\begin{center}
\begin{tabular}{lcccc}
\hline
Source    & V$_{\rm sys}$ & W$_{50}$ & Integrated Flux density      & \hi\ mass\\
          & \kms\        & \kms\    & Jy~\kms\  &  $10^7$~\msun\\
\hline
1         & $1907\pm  5$ & $ 29.5\pm10.5$ & $0.09\pm0.02$ & $1.6\pm0.3$\\
2         & $1930\pm 16$ & $166.5\pm38.6$ & $0.38\pm0.04$ & $6.9\pm0.9$\\
3         & $1966\pm  2$ & $ 45.1\pm3.8$  & $2.15\pm0.09$ & $39.7\pm1.5$\\
4         & $1990\pm  5$ & $135.5\pm10.8$ & $1.47\pm0.06$ & $27.2\pm1.0$\\
\hline
\end{tabular}
\end{center}
\label{tab:vflux}
\end{table*}

\section{Discussion}
\label{sec4}

\begin{figure}
\includegraphics[width=7.0cm]{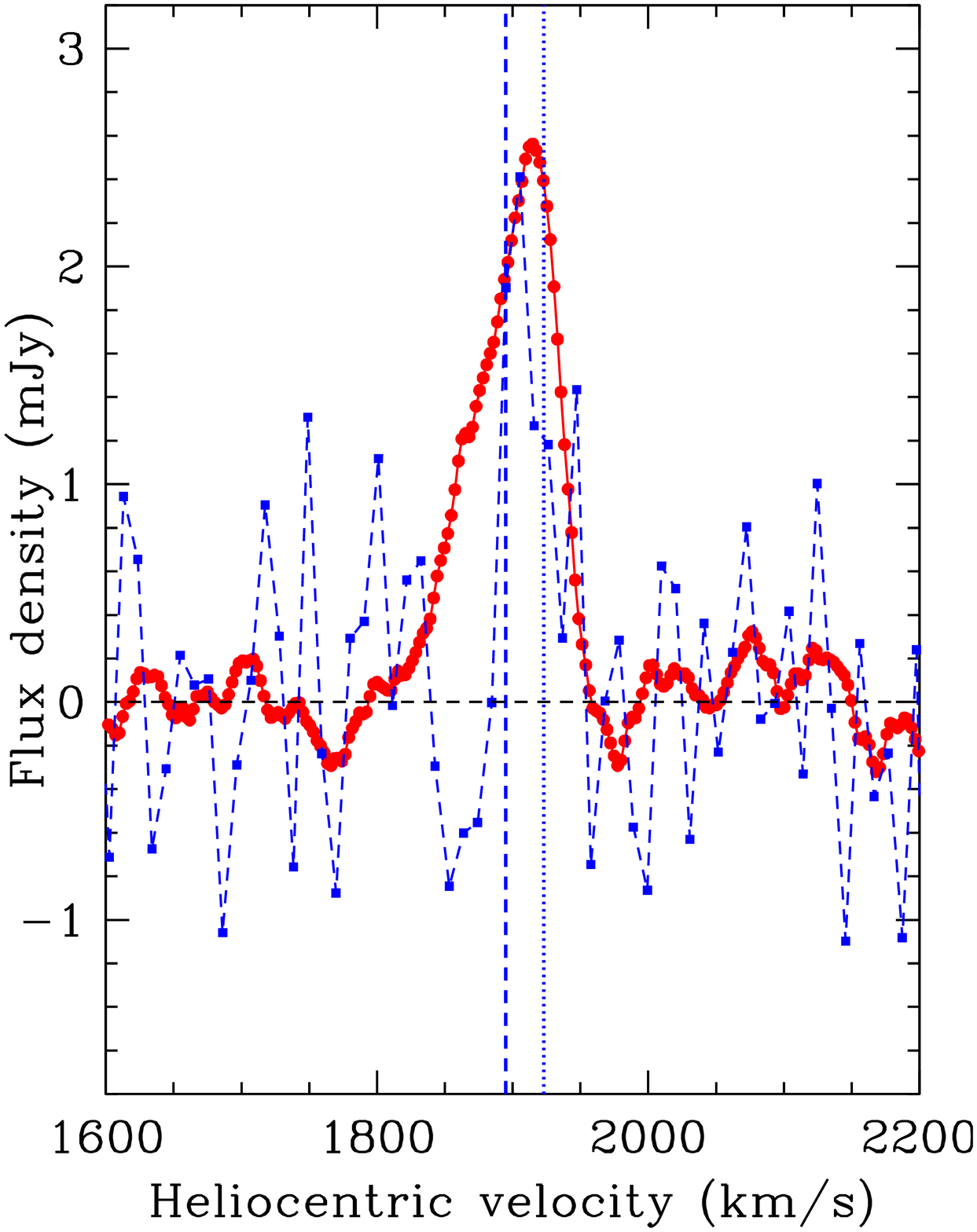}
\caption{A comparison of the Arecibo \hii\ spectrum derived from the central pointing 
towards PG\,1216+069 (in solid circles and the solid line) and the \hii\ spectrum 
derived from the VLA spectral cube (in solid squares, and the dashed line). The 
channel spacings of the two spectra are $\approx 24$~\kms\ (Arecibo) and 
$\approx 10.4$~\kms\ (VLA). The extended tail of \hii\ emission 
seen towards lower  velocities at Arecibo does 
not appear to be detected at the VLA. The dashed vertical line indicates the redshift 
measured from the metal lines (viz. $z = 0.00632$), while the dotted line indicates 
the velocity of the weaker absorption seen in the metal lines by \citet{tripp05}. 
See the main text for more details.
}
\label{fig:arecibo_vla}
\end{figure}

The \hii\ flux density detected in the central Arecibo pointing can be used 
to infer the \hi\ mass of the DLA host; this yields an \hi\ mass of 
$\sim 3.2 \times 10^7$~\msun. As can be seen from Fig.~\ref{fig:mom0a}, the 
HPBW of the Arecibo pointing centred on PG\,1216+069 includes \hii\ emission 
from Source~1 and part of Source~2. Fig.~\ref{fig:arecibo_vla} shows the 
Arecibo \hii\ spectrum (smoothed to a resolution of 24~\kms) along with the 
\hii\ spectrum derived from the VLA observations. It is apparent that the VLA \hii\ 
profile does not contain the tail towards lower velocities that is seen in the 
Arecibo spectrum. It should be emphasized that Source~2 is at a higher velocity 
than Source~1 (see Table~\ref{tab:vflux}); thus, even if part of the emission 
from Source~2 contributes to the Arecibo spectrum, it will not reduce the above 
discrepancy. It thus seems clear that there is some diffuse \hii\ emission at 
lower velocities that is detected at Arecibo, but not at the VLA. A comparison 
of the flux densities listed in Tables~\ref{5pt} and \ref{tab:vflux} indicates 
that  about 50\% of the flux density detected in the Arecibo \hii\ spectrum is 
missing from the VLA profile. In this context, it is worth noting that the \hii\ 
emission that we detect for VCC\,297 (i.e.  Source~4) from the VLA cube is in 
excellent agreement with that detected in an earlier Arecibo study \citep{giovanelli97}, 
indicating that, in general, the VLA observations are not systematically missing flux. 

The results from the multiple Arecibo pointings indicate that the \hii\ emission detected 
at Arecibo is well centred on the quasar PG\,1216+069. The measured \hii\ flux 
densities in the different pointings are consistent with the emitting region being 
considerably smaller than the Arecibo HPBW ($\approx 3.4^{'}$ at the line frequency 
of 1410~MHz). For example, a Gaussian emitting region of HPBW $1^{'}$ would only broaden 
the beam by $\approx 4$\%, while an HPBW of $2^{'}$ would broaden it by $\approx 16$\%. 
The slightly larger average normalised values of the \hii\ line intensity in the 
north-south direction are expected as the Arecibo beam is mildly elliptical (axial 
ratio~$\sim 1.15$), with the maximum lying along the direction of constant azimuth.  
Little evidence for systematic rotation can be discerned from the penultimate column of 
Table~\ref{5pt}, implying that the system is either (a)~face-on (if disk-like),
(b)~much smaller than $3.4^{'}$, or (c)~has no ordered rotation.

There have been several previous observations that attempted to detect \hii\ 
emission from the $z = 0.00632$ sub-DLA. \citet{kanekar05c} used the Giant
Metrewave Radio Telescope (GMRT) to obtain a $5\sigma$ limit of $0.06$~Jy\,\kms\
on the integrated \hii\ flux density (assuming a velocity width of 20~\kms),
thus obtaining the $5\sigma$ upper limit of $\sim 1.1\times 10^{7}$~\msun\ 
on the \hi\ mass of the absorber. Later, \citet{briggs06} used the WSRT to obtain 
a tentative detection of \hii\ emission, estimating the \hi\ mass to lie in the range 
$\sim (0.5 - 1.5)\times 10^{7}$~\msun\
(consistent with the GMRT non-detection), with a line peak at a velocity of 
$\sim 1905$~\kms. This is in agreement with the \hii\ emission detected in
our VLA spectrum, for which the measured flux density of 
Table~\ref{tab:vflux} implies an \hi\ mass of $(1.5\pm0.3) \times 10^7$~\msun.
Unfortunately, the WSRT beam is highly elongated at the low declination of 
the field, making it difficult to determine the morphology of the gas 
associated with the sub-DLA from the WSRT image. However, \citet{briggs06} 
note that there is a hint that the \hii\ emission is extended towards the 
north-east. \citet{briggs06} also detected \hii\ emission from the 
galaxies VCC\,297 and VCC\,415 (which lies outside the velocity range 
covered by the VLA observations), as well as the source labelled 3 in 
Fig.~\ref{fig:mom0a}. We note that \citet{tripp08} report an \hi\ mass 
of $\sim 8 \times 10^6$~\msun\ for the host galaxy of the sub-DLA (which 
corresponds to an \hi\ mass of $\sim 6 \times 10^6$~\msun\ for the distance 
adopted here). This is smaller (at $\sim 3\sigma$ significance) than our \hi\ 
mass estimate. \citet{tripp08} do not give any details of their data analysis, 
and the reasons for this discrepancy are unclear.

From Fig.~\ref{fig1}, the peak \hii\ brightness of the sub-DLA is $2.8\,\mjyb$. 
The heliocentric velocity at the peak \hii\ intensity is $1917.6 \pm 3.4$~\kms, 
corresponding to a heliocentric redshift of $z=0.006396$. The line profile is 
seen to be asymmetric, extending to lower velocities on the blue side. The line 
velocity width at half power is $68 \pm 3.4$~\kms, while the velocity width at 
zero intensity is far wider, $150 \pm 17$~\kms. Interestingly, the weaker 
component of the low-ionization metal absorption lines detected in the 
sub-DLA \citep{tripp05} lies at the peak velocity of the \hii\ emission, while 
the stronger metal absorption component lies in the extended blue wing of the 
\hii\ line.  Fig.~\ref{fig1} also shows a two-Gaussian fit to the Arecibo 
profile. The parameters (amplitude, position and standard deviation) of the 
Gaussians are $2.1 \pm 0.5 \times 10^{-3}$~Jy, $1921.2 \pm 1$~\kms, and $13.3 \pm 2$~\kms, 
and  $1.5 \pm 0.2 \times 10^{-3}$~Jy, $1885.9 \pm 8$~\kms, and $27.2 \pm 5$~\kms. 
If we assume that the broad component arises from a diffuse ($\sim 2^{'}$-sized) 
cloud, the resulting flux per beam falls below the RMS noise ($\sim 0.4$~mJy for 
a velocity resolution of $\sim 20$~\kms\ and a spatial resolution of $\sim 63^{''}$)
of the VLA data cube, consistent with this emission not being detected 
in the VLA observations.

What is the nature of the \hi\ clouds that give rise to the observed \hii\ emission?
For our assumed distance of 28~Mpc, the absolute blue magnitude of the nearby 
galaxy VCC\,297 is $-16.6$~mag. \citep{gavazzi06}. If the ratio of the \hi\ mass 
to the blue luminosity in Source~1 is the same as that in VCC\,297, one would 
expect it to have $\mb \sim -13.2$~mag., or an apparent magnitude of 18.2~mag. 
The corresponding numbers for Source~2 and Source~3 are $\mb\sim -15.1$~mag.
and $\mb\sim -17.0$~mag., respectively, or apparent magnitudes of 16.7~mag. and
14.8~mag., respectively. Note that since we are scaling from the values
for VCC\,297, the apparent magnitudes that we estimate are independent
of the assumed distance. Deep optical images of the PG\,1216+069 field have been 
obtained by \citet{prochaska11} and \citet{tripp08}. Unfortunately, the angular 
resolution of the VLA \hii\ image is relatively poor, $\approx 1'$, and there are hence
multiple optically-identified galaxies within a distance of $0.5'$ of the 
individual \hi\ clouds. However, none of these galaxies are as bright as the 
scaled estimates from VCC\,297. For example, the brightest objects within 
$0.5'$ of Sources~2 and 3 have B- and g-magnitudes of $\approx 21.8$ 
\citep{prochaska11} and $\approx 21.0$ (SDSS), respectively, $\approx 5$~mag. fainter 
than the scaled estimates from VCC\,297. Thus, there appears to be little star formation
associated with the neutral gas in Sources~2 and 3. Consistent with this, 
\citet{gavazzi12} do not detect H-$\alpha$ emission from Source~3, obtaining an 
upper limit of SFR~$< 0.0005$~M$_\odot$~yr$^{-1}$. The faintness of the optical
counterparts of Sources~2 and 3 thus rules out the possibility that these \hi\ clouds 
arise in galaxies similar to VCC\,297.

In the case of Source~1, its proximity to the background quasar \citep[V magnitude of 
$\sim 15.4$~mag;][]{hamilton08} implies that it would be difficult to detect a faint 
dwarf galaxy of the above estimated absolute blue magnitude
\citep[even with HST imaging of the field;][]{chen01}. Indeed, the estimated
blue magnitude is similar to that measured for the post-starburst dwarf galaxy 
J1229+02 \citep{stocke04b} that has been suggested to
be associated with the metal absorption seen at a heliocentric velocity of 1585~\kms\
towards 3C273.  This system also lies in the outskirts of the Virgo cluster and 
the galaxy itself appears to be at an intermediate stage in the fading dwarf
evolutionary sequence proposed by \citet{babul92}. \citet{keeney14} suggest that 
the observed metal-line absorption system arises from galactic winds from this 
galaxy interacting with ambient material. 

The dynamical mass of the sub-DLA can be estimated as M$_{\rm dyn} \sim 
(\Delta V^2 \times R)/G$, where $\Delta V$ is the characteristic velocity 
and $R$, the characteristic size. If we take $\Delta V$ to be half of the FWHM, 
i.e. $\sim 34$~\kms, and $2R$ to correspond to $\sim 2^{'}$, i.e. intermediate 
between the VLA and the Arecibo resolutions, the dynamical mass is 
$\sim 10^9$~\msun, significantly larger than the estimated 
\hi\ mass of $3.2 \times 10^7$~\msun. Conversely, if we assume the size to be
that of the VLA resolution of $\sim 1^{'}$ then the dynamical mass would 
be $\sim 5\times 10^8$~\msun. Interestingly, if the size of the sub-DLA
is indeed $\sim 2^{'}$, our measured \hi\ mass corresponds to an average \hi\
column density of $\log(\nhi/{\rm cm}^{-2}) = 19.29$, very close to the measured value
of 19.32 along the quasar sightline \citep[][]{tripp05}. If this is indeed 
the typical \hi\ column density of the system, then it is not surprising 
that there is no associated star formation.

\citet{tripp05} measured an extremely low metallicity, [O/H]~$= -1.60^{+0.09}_{-0.11}$,
for the $z = 0.00632$ sub-DLA, from the O{\sc i}$\lambda$1302 line, with similar 
values obtained from multiple Si{\sc ii} lines. Their detection of the Fe{\sc ii} 
lines showed no evidence for the depletion of iron, indicating that the absorber 
contains very little dust. They also found nitrogen to be under-abundant in the 
sub-DLA, indicating that the absorber is a primitive, chemically-young galaxy
\citep{tripp05}. A preliminary analysis of a new HST {\it Cosmic Origins Spectrograph} 
ultraviolet spectrum of PG\,1216+069 (with a higher S/N ratio than that of the 
HST-STIS spectrum of \citealt{tripp05}) confirms both the very low metallicity and 
the underabundance of nitrogen reported by \citet{tripp05}.

While the \hi\ mass obtained for the sub-DLA towards PG\,1216+069 is typical of 
faint dwarfs, its very low metallicity, $\sim 1/40$ solar \citep{tripp05}, 
is very unusual in the local Universe. From the 
mass-metallicity relation for dwarf galaxies \citep{lee06}, the expected 
stellar mass for a dwarf galaxy with this metallicity is 
$\sim 5.4\times 10^4$~\msun, i.e. nearly 3 orders of magnitude lower than 
the \hi\ mass inferred from our Arecibo spectrum. However, we note that there 
are a few extreme dwarf galaxies which deviate significantly from the 
mass-metallicity relation at the low-metallicity end \cite[e.g.][]{ekta10}. 
For example, SBS0335-052W has a metallicity of $\sim 1/35$ solar and 
an \hi\ mass of $5.8\times10^8$~\msun\ \citep{izotov05,ekta09}. Further,
most of the metallicity measurements available in the literature refer
to star-forming regions in the centres of dwarf galaxies. Some dwarf
irregular galaxies do appear to have significant metallicity gradients
\citep[e.g.][]{pilyugin15}, and it is possible that gas stripped from the outer 
parts of such dwarfs could have a low metallicity.

\begin{figure}
\includegraphics[width=7cm]{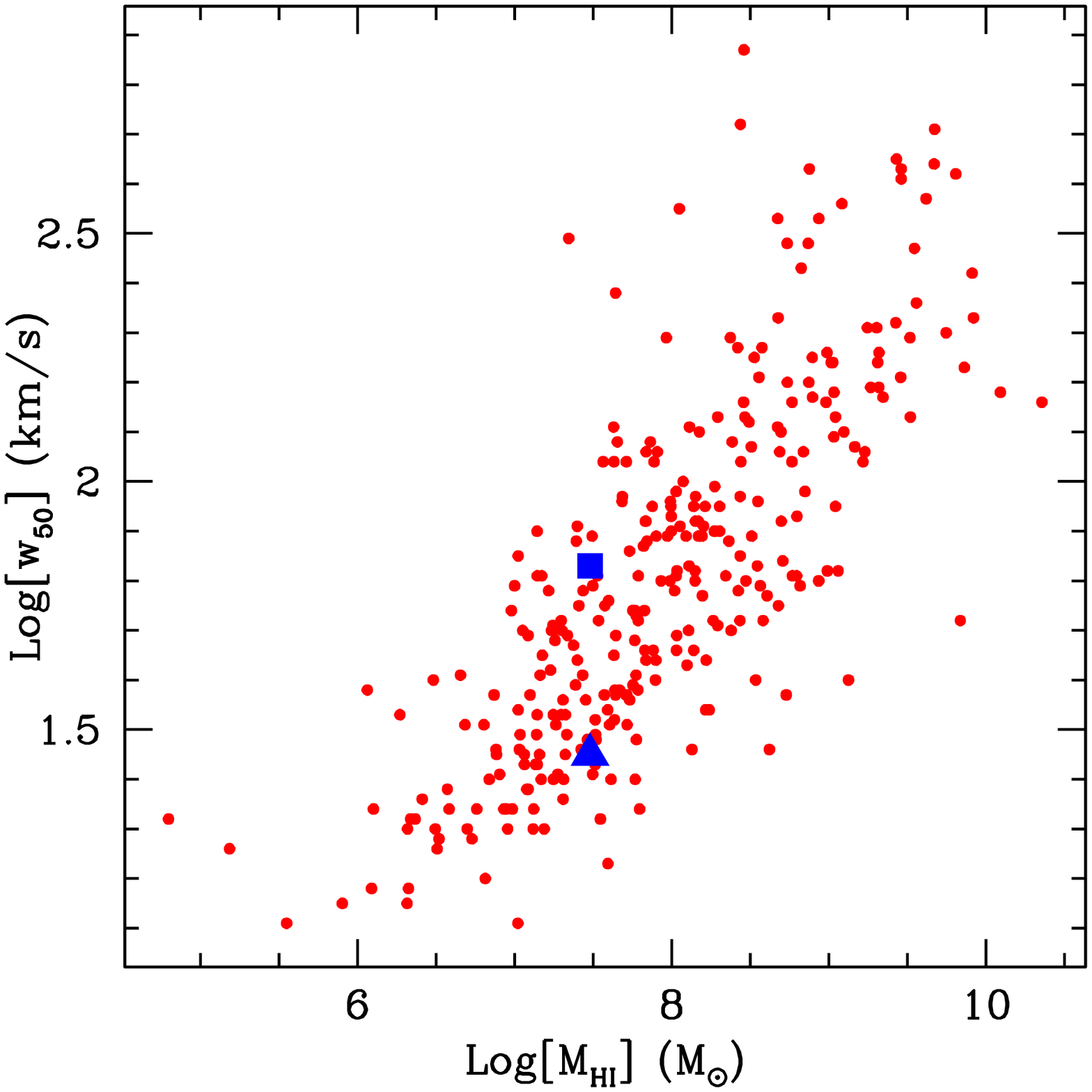}
\caption{The velocity width between 50\% points (W$_{50}$) plotted versus 
\hi\ mass for the galaxies in the \citet{karachentsev04} sample. With
an \hi\ mass of $\sim 3.2\times 10^7$~\msun and  W$_{50} \sim 68$~\kms, the
properties of Source-1 (indicated by the solid square) are not significantly 
discrepant from those of the galaxies in this sample. The solid triangle marks the
same \hi\ mass, but the velocity width is taken to be 28~\kms, corresponding
to the velocity separation between the two metal-line absorption systems 
associated with this system \citep{tripp05}. See the main text for more details.
}
\label{fig:w50}
\end{figure}

The velocity spread and asymmetry of the \hii\ emission are both interesting in 
terms of interpreting the nature of the host galaxy. Fig.~\ref{fig:w50} shows 
a plot of W$_{50}$ versus M$_{\rm HI}$ for galaxies from the Local Volume 
catalogue of \citet{karachentsev04}. At M$_{\rm HI} = 3.2 \times 10^{7}$ M$_{\odot}$,
a velocity width of W$_{50}$ of $\sim 68$~\kms\ appears reasonable. However, the 
highly asymmetric \hii\ line profile and the large ($\sim 150$~\kms) 
velocity separation between nulls are both very unusual for a dwarf galaxy
\citep[e.g.][]{begum08}. Similarly, the presence of significant amounts of 
diffuse gas (as evidenced by the discrepancy between the VLA and Arecibo 
\hii\ flux densities) is unusual for a dwarf galaxy. Thus, although we cannot 
definitively rule out the possibility that Source~1 is a dwarf galaxy, the 
available evidence suggests that, like Sources~2 and 3, it too is a hydrogen 
cloud with very little associated star formation.

\begin{figure}
\includegraphics[width=7.0cm]{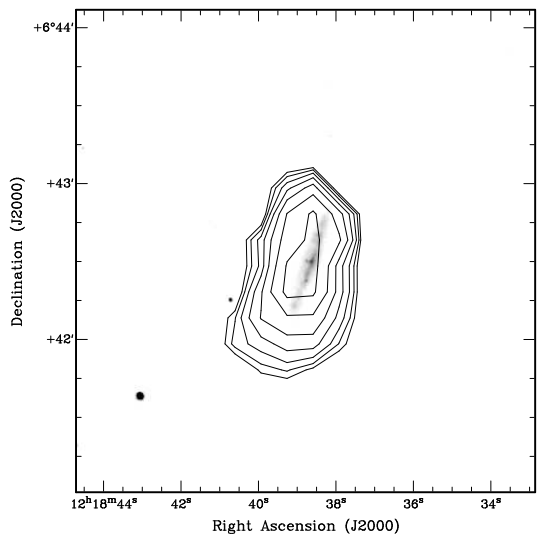}
\caption{A higher resolution ($39^{''}\times 39^{''}$) VLA image of the \hii\ 
emission from VCC297 (contours) overlayed on the Sloan Digital Sky Survey (SDSS) 
g-band image (greyscale). The \hii\ contours start at  $2 \times 10^{19}$ atoms~cm$^{-2}$
and increase in steps of $\sqrt(2)$. The \hii\ emission appears to be extended 
towards the southern side of the galaxy, possibly because of ram pressure
from the surrounding medium. Two of the other sources in this field
(Sources~2 and 3; see Fig.~\ref{fig:mom0a}) also appear to be
extended in the north-south direction.
}
\label{fig:vcc297}
\end{figure}

Several such ``dark'' \citep[and ``almost dark''; ][]{cannon15} \hi\ clouds 
are known near the Virgo cluster, including HI\,1225+01 \citep{giovanelli89,chengalur95}, 
Virgo\,HI21 \citep{minchin05,minchin07}, and several objects discovered as part of the 
ALFALFA survey \citep{kent07}. The origin of these clouds is as yet unclear. For example, 
in the case of Virgo\,HI21, which has an \hi\ mass of $\sim 10^8$~\msun, the proposed 
origins include (1)~a dark galaxy \citep{minchin05}, and (2)~a part of a long ($\sim 250$~kpc) \hi\ 
tail stripped from NGC\,4254 as it falls at high speed into the Virgo cluster 
\citep{haynes07}. However, the field under study here is further unusual in 
that it contains 4 \hi\ sources, of which 3 have no optical counterpart. 
Similarly, \citet{janowiecki15} discuss a system of three objects (once again lying 
behind the Virgo cluster), where only the object with the highest \hi\ flux has an 
optical counterpart. The origin of the other two \hi\ clouds is unclear. The Virgo 
cluster appears to be a rich hunting ground for such small groups of clouds: 
\cite{kent09} discuss a field in which there are 5 \hi\ clouds. Interferometric studies 
of the \hi\ clouds in the Virgo cluster have clearly identified  some of these 
objects as debris from larger galaxies, but the origin of several other clouds 
remains unclear \citep[e.g.][]{kent09,kent10}. \citet{kent09} present a model in 
which the clouds have been stripped from a larger galaxy on its first passage through
the intra-cluster medium (ICM), estimating  the evaporation timescale 
(due to interactions with the hot ICM) for such clouds to be very short, 
$\lesssim 10^8$~yr. They find that such clouds could be present at 
distances as large as $\approx 240$~kpc from the parent galaxy. 
Other studies have also demonstrated the importance of ram pressure 
stripping of galaxies falling into the  Virgo cluster. For example, NGC\,4388 has a 
$\sim 100$~kpc-long tail that is believed to have been produced by ram pressure stripping 
\citep{oosterloo05}. \cite{chung07} find that several of the spiral galaxies with 
projected separations of $< 1$~Mpc from M87 have \hi\ tails pointing away from M87, 
again indicative of gas stripping due to ram pressure. In this context, it is 
interesting to note that both  Sources~2 and 3 are extended in the north-south 
direction (i.e. roughly radially from M87). Fig.~\ref{fig:vcc297} shows a higher 
resolution VLA image of the integrated \hii\ emission of VCC297; the extension of 
the emission in the north-south direction is again quite clear. Unfortunately, at this 
resolution, no emission from Sources~1 and 2 is detected, and only the dense clump 
of Source~3 is seen. The fact that the majority of the sources detected in this field 
are extended along the direction towards M87 suggests that these gas clouds may be 
debris left behind due to ram-pressure stripping of a small 
group of galaxies that is falling into the Virgo cluster. The total mass of the clouds 
that we detect (i.e. Sources~1, 2 and 3) is  $\sim few \times 10^{8}$~\msun, similar 
to the total mass of the cloud  complex studied by \citet{kent09}. As discussed in 
\citet{briggs06}, there are at least 6 known galaxies within 400~\kms\ of Source~1 
located in a square region of size $\approx 300$~kpc around the source. 
In principle, any of these galaxies could be the source of the clouds that we 
detect in the VLA image. However, the entire cloud complex is located $\gtrsim 2$~Mpc 
in projection away from M87 \citep[although it does lie projected on the outskirts of 
the X-ray emission around M49;][]{bohringer94} and, if we assume it to be part of 
complex ``W'' at twice the Virgo distance, it is unclear whether the ram pressure 
at this location would be sufficient to strip the gas from the parent
galaxy. Further, even for the sources that are extended in the north-south
direction, the tails themselves point in different directions. An even
more serious consideration is the  extremely low metallicity of the gas detected 
in absorption in Source~1. As discussed above, there are very few known 
galaxies in the local Universe with such low metallicities; it hence appears 
unlikely that this gas arises from the stripping of the interstellar medium 
of a normal galaxy. Thus, although the currently available observations have 
significantly advanced our understanding of the host of the $z = 0.00632$ 
sub-DLA, they also raise several further puzzles.

\section{Summary}

We present the results of Arecibo observations of \hii\ emission from 
the $z \sim 0.00632$ sub-DLA seen towards the quasar PG\,1216+069. 
\hii\ emission is clearly detected at Arecibo, yielding an \hi\ mass of
$\sim 3.2\times 10^7$~\msun. From a mosaic of pointings around the quasar 
location, we find that the \hii\ emission is not extended in comparison to the
Arecibo angular resolution of $\sim 3.4^{'}$. The offset pointings also show 
no evidence for a systematic velocity gradient in the \hii\ emission. 

Our analysis of archival VLA data on this source also yields a detection of 
\hii\ emission in images with an angular resolution of $\sim 63^{''}$. However,
the VLA images recover only about 50\% of the flux detected at Arecibo.
We also detect \hii\ emission from 3 other sources in the same field of view, 
only one of which is associated with a known optically-detected galaxy. We find 
the \hii\ emission from all of the sources, except that which hosts the 
sub-DLA (for which the S/N is very low), to be extended in the north-south direction,
although the emission tails are themselves not aligned.  The entire complex 
appears to be part of complex ``W'' that is believed to be behind the
Virgo cluster, and at about twice its distance. It is unclear whether the 
gas stripping mechanisms that operate in clusters are relevant at the large
distance ($\gtrsim 2$~Mpc) of this complex from Virgo. Further, the extremely low 
metallicity of the sub-DLA ($\sim 1/40$ solar) makes it unlikely that this material 
has been stripped from a normal galaxy; there are only a handful of galaxies with 
such low metallicities in the local Universe. In summary, while the $z \sim 0.006$ 
absorber towards PG\,1216+-69 is one of only two known DLAs or sub-DLAs with 
detected \hii\ emission, providing an estimate of its total gas mass, there 
are several unresolved puzzles about the origin of this system.

\section{Acknowledgments}

The Arecibo Observatory is operated by SRI International under a cooperative 
agreement  with the National Science Foundation (AST-1100968), and in alliance
with Ana G. M{\'e}ndez-Universidad Metropolitana, and the Universities Space
Research Association. The National Radio Astronomy Observatory is a facility
of the National Science Foundation, operated under cooperative agreement by 
Associated Universities, Inc. NK acknowledges support from the DST, India, 
via a Swarnajayanti Fellowship.  We are grateful to the anonymous referee 
for comments which have led to improvements in the paper.

\bibliographystyle{mn2e}
\bibliography{ms}

\end{document}